\documentclass{ws-procs9x6}

\begin{document}
\title{BPS Wall Solutions in Five Dimensional Supergravity
\footnote{\uppercase{T}alk presented by \uppercase{M. Arai}
at {\it \uppercase{SUSY} 2003:
\uppercase{S}upersymmetry in the \uppercase{D}esert}\/, 
held at the \uppercase{U}niversity of \uppercase{A}rizona,
\uppercase{T}ucson, \uppercase{AZ}, \uppercase{J}une 5-10, 2003.
\uppercase{T}o appear in the \uppercase{P}roceedings.}}

\author{Masato Arai}

\address{Institute of Physics, AS CR, 
  182 21, Praha 8, Czech Republic}

\author{Shigeo Fujita, Masashi Naganuma, Norisuke Sakai}

\address{Department of Physics, \\
  Tokyo Institute of Technology \\
  Tokyo 152-8551, JAPAN}


\maketitle

\abstracts{We find 
an exact solution of BPS wall in five-dimensional 
supergravity using a gravitational 
deformation of the massive Eguchi-Hanson nonlinear sigma model. 
The warp factor decreases for both infinities of the extra 
dimension. 
Our solution requires no 
fine-tuning between boundary and bulk 
cosmological constants, in contrast to 
the Randall-Sundrum model. 
Wall solutions are also obtained with 
warp factors which are flat or increasing in 
one side 
by varying a deformation parameter. 
}

\section{Introduction} 
Randall and Sundrum have proposed 
one of the most interesting models 
in the brane-world scenario, which exhibits 
the localization of four-dimensional 
 graviton \cite{RS2} 
 by a metric 
 containing a warp factor $e^{2U(y)}$ which decreases 
 exponentially for both infinities of the extra dimension 
$y \rightarrow \pm \infty$ 
\begin{eqnarray}
  ds^2=
g_{\mu \nu}dx^\mu dx^\nu = e^{2U(y)}\eta_{mn}dx^mdx^n + dy^2,
\label{5dmetric}
\end{eqnarray}
where $\mu, \nu = 0,..,4$, $m, n = 0,1,3,4$ and $y\equiv x^2$. 
A bulk cosmological constant 
 and a boundary cosmological constant had to be introduced 
 and fine-tuned each other. 

This scenario uses an orbifold which may be 
regarded as a delta-function like 
domain wall.
It is desirable to obtain the domain wall as a classical
 solution in some field theory from 
 a phenomenological point of view.
It has been shown that 
 domain wall solutions in gauged 
 supergravity theories 
 in five dimensions require hypermultiplets 
\cite{KalloshLinde} to obtain 
 warp factors decreasing for both 
 infinities $y \rightarrow \pm \infty$ 
 (infra-red (IR) fixed points in AdS/CFT correspondence). 
The target space of 
hypermultiplets in 
 five-dimensional supergravity theory must be 
 quaternionic K\"ahler (QK)
 manifolds \cite{BaggerWitten}.
Domain walls in massive QK nonlinear sigma model 
(NLSM) in supergravity
 theories have been studied using homogeneous \cite{Alekseevsky,BC2} and
 inhomogeneous manifolds \cite{Beh-Dall,Lazaroiu}. 
Warp factor in the latter models connects between 
IR fixed points, as is desirable phenomenologically.
However, these manifolds do not allow a limit of weak 
 gravitational coupling.

The purpose of this paper is to give an exact BPS domain wall 
 solution in five-dimensional supergravity 
 coupled with hypermultiplets 
 (and vector multiplets).
Our strategy to construct the model is to deform the NLSM 
 in SUSY theory having domain wall solution 
 to the model with gravity.
Massive hyper-K\"ahler NSLMs without gravity 
 in four dimensions have been constructed 
 in harmonic superspace as well as {\it N}$=1$ superfield
 formulation \cite{ANNS}, 
 and have yielded the domain wall 
 solution for the Eguchi-Hanson (EH) manifold.
Inspired by this solution,
 we deform this model into five-dimensional supergravity model
 and we consider the BPS domain wall solution.
This paper is based on our original paper \cite{ANFS}.

\section{Bosonic action of our model in 5D Supergravity}
To find a gravitational deformation of the NLSM
 with EH target manifold, 
 we use the off-shell formulation of five-dimensional 
 supergravity
 \cite{Fujita-Ohashi,FKO} 
 combined with the quotient method via a vector multiplet without 
 kinetic term and the massive deformation.
We start with the system of a Weyl multiplet, three hypermultiplets 
 and two $U(1)$ vector multiplets. 
One of the two 
 vector multiplets has no kinetic term and plays the role of a Lagrange 
 multiplier for hypermultiplets to obtain a curved target manifold. 
The other vector multiplet, which is referred to as $U(1)_0$ vector
 multiplet in the following, serves 
 to give mass terms for hypermultiplets. 
The deformation parameters are the gravitational coupling constant
 $\kappa$ and $a$ being a constant in generator of 
 $U(1)_0$ gauge symmetry.

After integrating out the auxiliary fields 
 by their on-shell conditions in the off-shell supergravity
 action, we obtain the bosonic part of the action 
 for our model with two kinds of constraints.
One of constraints comes from the gauge fixing 
 of dilatation, and 
 makes target space of hypermultiplets be a non-compact version of 
 quaternionic projective space, $\frac{Sp(2,1)}{Sp(2)\times Sp(1)}$, 
 combined with the gauge fixing of 
 $SU(2)_R$ symmetry. 
The other is required by the on-shell condition 
 of auxiliary fields of the $U(1)$ vector multiplet without kinetic term, 
 and corresponds to the constraint for the EH target space 
 in the limit of $\kappa\to 0$.

After solving these constraints, bosonic part
 of the Lagrangian can be described in terms of
 independent four real scalar fields $(r,\theta,\Psi,\Phi)$ 
 (eight remnant scalars are eliminated) besides the graviton.
The scalar potential can be described by $r$ and $\theta$,
 and it is found that the theory has two discrete local vacua 
 at $(r,\theta)=(0,0),(0,\pi)$ in small $\kappa$. 
We can thus expect domain wall solution connecting them. 
Note that these local vacua become saddle points 
 as $\kappa$ increases.
We consider the case of small $\kappa$ in what follows.

\section{BPS equation and the solution} 
Instead of solving Einstein equations directly, we solve BPS equations  
 to obtain a classical solution conserving a half of SUSY. 
Since we consider bosonic configurations, we need to examine 
 the on-shell SUSY transformation of gravitino and hyperino.
The condition to preserve four SUSY is specified by 
$
  \gamma^y \varepsilon^i(y) = i \tau_3^{i}{}_j \varepsilon^j(y),
$
where $\tau_3$ is one of the Pauli matrix.
Substituting this condition and the metric ansatz (\ref{5dmetric}) 
 into the on-shell SUSY transformation of gravitino and hyperino,
 we obtain BPS equation.
Solving the BPS equation for hyperino,
 the wall solution interpolating between the two vacua
 $(r, \theta)=(0,0),(0,\pi)$ is obtained as
$
 r=0,~\cos \theta = \tanh \left(2g_0M^0(y-y_0)\right),~ 
\Phi=\varphi_0,
$
with $\Psi$ undetermined, 
 where $g_0$ and $M^0$ 
 are $U(1)_0$ coupling constant and scalar of 
 $U(1)_0$ vector multiplet fixed as
 $M_0=\sqrt{3/2}\kappa$, respectively,
 and $y_0$ and $\varphi_0$ are constants.
Here we take the boundary condition $r=0$ at $y=-\infty$.
Using this solution,
 we obtain the BPS solution of the warp factor and the Killing 
 spinor from SUSY transformation for gravitino as
\begin{eqnarray}
 U(y) = -\frac{\kappa^2\Lambda^3}{3(1-\kappa^2\Lambda^3)}
        \left[ \ln \{\cosh \left(2g_0M^0(y-y_0)\right)\} 
           + 2ag_0M^0(y-y_0)\right],
\end{eqnarray}
and 
$
 \varepsilon^i(y)\equiv e^{U(y)/2}\tilde{\varepsilon}^i,
 (\gamma^y \tilde{\varepsilon}^i = i\tau_{3}^{i}{}_j\tilde{\varepsilon}^j).
$
Here $\Lambda$ is a constant having mass dimension one
 and $\tilde{\varepsilon}^i$ is a constant spinor.

The warp factor $e^{2U(y)}$ of this solution 
 decreases exponentially for both infinities $y\rightarrow \pm \infty$ 
 for $|a|<1$.
In this case, BPS wall solutions interpolate two IR fixed points
 in boundary field theories. 
The case of
 $|a|=1$ becomes the wall solutions interpolating 
 between AdS and flat Minkowski vacua. 
On the other hand, warp factor increases exponentially either one of the 
 infinities for $|a|>1$.
The wall solutions interpolate one IR and one UV fixed points. 
The family of our BPS solutions contains a parameter $a$ interpolating 
 between these three classes of field theories.

We find that our BPS solution for $a=0$ 
 automatically satisfies the fine-tuning condition. 
By taking $g_0M^0 \rightarrow \infty$ 
 and $\Lambda \rightarrow 0$ with $g_0M^0\Lambda^3$, $\kappa$, 
 and $a$ fixed, we can obtain thin wall limit. 
Substituting the BPS wall solutions
 into the Lagrangian and taking the thin wall limit, 
 we obtain wall tension and bulk cosmological constant as 
 $T_w = 4(g_0M^0\Lambda^3)$ and 
 $\Lambda_c 
 = -\frac{8\kappa^2(g_0M^0\Lambda^3)^2}{3}$,
 respectively.
These satisfy the fine-tuning condition 
 of the Randall-Sundrum model
 $\sqrt{-\Lambda_c}=\frac{\kappa}{\sqrt{6}}T_w$. 
Therefore we have realized the single-wall Randall-Sundrum model  
 as a thin-wall limit of our solution of the coupled scalar-gravity theory, 
 instead of an artificial boundary cosmological constant put 
 at an orbifold point. 

\section*{Acknowledgements}
M.~A. is grateful to the organizers of SUSY2003.
This work is supported in part by Grant-in-Aid 
 for Scientific Research from the Japan Ministry 
 of Education, Science and Culture  13640269 (NS). 

\end{document}